\documentclass[fleqn, usenatbib]{mnras}
\usepackage[T1]{fontenc}
\usepackage{ae,aecompl}
\usepackage{graphicx}
\usepackage{color}
\usepackage{amsmath}

\title[Schwarzschild method and non-sphericity]{The effect of non-sphericity on mass and anisotropy measurements
in dSph galaxies with Schwarzschild method}

\author[K. Kowalczyk et al.]{
Klaudia~Kowalczyk$^{1}$\thanks{E-mail: \href{mailto:klaudia.kowalczyk@gmail.com}{klaudia.kowalczyk@gmail.com}},
Ewa L.~{\L}okas$^{1}$
and Monica~Valluri$^{2}$
\\
$^{1}$Nicolaus Copernicus Astronomical Center, Polish Academy of Sciences, Bartycka 18, 00--716 Warsaw, Poland\\
$^{2}$Department of Astronomy, University of Michigan, 1085 South University Ave., Ann Arbor, MI 48109, USA
}

\pubyear{2018}

\begin{document}
\label{firstpage}
\pagerange{\pageref{firstpage}--\pageref{lastpage}}
\maketitle

\begin{abstract}
In our previous work we confirmed the reliability of the spherically symmetric Schwarzschild orbit-superposition
method to recover the mass and velocity anisotropy profiles of spherical dwarf galaxies. Here we investigate
the effect of its application to intrinsically non-spherical objects. For this purpose we use a model of a
dwarf spheroidal galaxy formed in a numerical simulation of a major merger of two disky dwarfs.
The shape of the stellar component of the merger remnant is
axisymmetric and prolate which allows us to identify and measure the bias caused by observing the
spheroidal galaxy along different directions, especially the longest and shortest principal axis.
The modelling is based on mock data generated from the remnant that are observationally available for dwarfs:
projected positions and line-of-sight velocities of the stars. In order to obtain
a reliable tool while keeping the number of parameters low we parametrize the total mass distribution as
a radius-dependent mass-to-light ratio with just two free parameters we aim to constrain.
Our study shows that if the total density profile is known, the true, radially increasing anisotropy profile
can be well recovered for the observations along the longest axis whereas the data along the shortest axis
lead to the inference an incorrect, isotropic model. On the other hand, if the density profile is derived from
the method as well, the anisotropy is always underestimated but the total mass profile is well recovered for the data
along the shortest axis whereas for the longest axis the mass content is overestimated.
\end{abstract}


\section{Introduction}

Galaxies in the observed Universe are divided into a few distinct morphological types based primarily on their shapes,
sizes, the nature of their stellar populations and gas content. Giant ellipticals are dominated by non-spherical
stellar components. While elliptical galaxies are expected to be embedded in a dark matter halo, the baryons (primarily
an old stellar population) tend to dominate the dynamics over most of the region where the stars are found. Spiral
galaxies are characterized by a thin stellar disk containing gas which is actively forming stars and older central
components which often include a stellar bulge, a stellar bar and a supermassive black hole. Although the dynamics of
gas and stars in the outer regions are strongly affected by dark matter, inner regions are still baryon dominated.
Baryons also dominate much of the dynamics for smaller dwarf elliptical and dwarf irregular galaxies.

For all these types of galaxies a significant fraction of the total mass in the central parts is contained in the visible
components: stars and gas. The situation is different for \textit{dwarf spheroidal} (dSph) and ultra-faint (UFD)
galaxies. Their high line-of-sight velocity dispersions cannot be explained within Newtonian dynamics without the
addition of heavy (when compared to the mass in stars) dark matter haloes. Estimated ratios between the masses of dark
and baryonic matter in dwarf galaxies reach hundreds (\citealt{mateo_1998}, \citealt{gilmore_2007}).
They are thought to have formed in the least massive haloes and during their evolution accreted or been able to retain
much less baryonic matter (\citealt{governato_2010}, \citealt{sawala_2016}).

While the mass enclosed within some characteristic radius can be determined with simple estimators, independent of the
orbit anisotropy (\citealt{walker_2009}, \citealt{wolf_2010}), detailed studies of mass distribution require more
sophisticated methods. The most widely used method is based on solving the spherical Jeans equation \citep{GD} but this
method suffers from the well known mass-velocity anisotropy degeneracy \citep{binney_1982} since the velocity
anisotropy cannot be determined directly. Significant improvements to the Jeans equation method have been made by
including kurtosis in a fitting procedure which partially lifts the mass-anisotropy degeneracy  \citep{lokas_2002,
lokas_2005}. However, the assumptions regarding the spherical symmetry and a particular form of the velocity
anisotropy profile are usually necessary.

A more general method that does not require such assumptions is the orbit superposition Schwarzschild modelling
\citep{schwarzschild_1979}. It has been successfully used in the last few decades in modelling ellipticals and bulges
with spherically symmetric, axisymmetric and triaxial codes (\citealt{vdMarel_1998}, \citealt{cretton_1999},
\citealt{gebhardt_2003}, \citealt{valluri_2004}, \citealt{thomas_2004}, \citealt{cretton_2004}, \citealt{cappellari_2006},
\citealt{vdBosch_2010}). Recently, it has also been applied to dwarf spheroidals, generally assuming sphericity of both
stars and dark matter halo, and focusing on studying the inner slope of the density profile of the halo
(\citealt{breddels_2013}, \citealt{breddels_2013b}, \citealt{jardel_2012}, \citealt{jardel_2013}).

Unfortunately, observed shapes of dwarf galaxies in the Local Group show non-negligible ellipticities
\citep{mcconnachie_2012}. It can be explained within a scenario in which dwarf galaxies were accreted by their hosts as
disky and require long evolution on a tight orbit to become spherical (\citealt{lokas_2012},
\citealt{kazantzidis_2013}). Non-spherical dwarfs are also the most natural outcome of mergers between disky dwarfs, as
we discuss below.
Note that both simulations of mergers and tidal evolution of rotationally supported dwarfs favor
prolate and triaxial shapes over oblate ones \citep{kazantzidis_2011a, kazantzidis_2011b}.
Nevertheless, a possible impact of the ellipticity should not be a reason to relinquish spherically
symmetric modelling methods. Their ostensible simplicity proves to be their strength when, as for dwarfs, available
data are limited and a low number of free parameters of a model is desired.

Such simplification, however, necessarily leads to systematic errors in the results. Therefore, when deciding to
proceed with a spherically symmetric method, it is very important to be aware of the existing biases and their
magnitude. They can be measured by applying the method to mock data obtained by observing similar objects formed in
numerical simulations along different lines of sight. Such studies have been done for the simple mass estimators
mentioned before. \citet{kowalczyk_2013} measured the bias caused by the triaxiality of the stellar component
and compared the masses estimated for simulated dwarfs as a function of the axis ratio. They showed that the mass is
underestimated when the line of sight towards the galaxy is aligned with the shortest axis of the stellar
component, fairly well recovered for the intermediate axis and overestimated by up to a factor of two for the longest
one. \citet{campbell_2017} applied a slightly different approach to more advanced simulations, averaging over many lines
of sight in order to derive the mean and proving that on average the estimators are unbiased.

In spite of applying the spherically symmetric Schwarzschild method to dwarf spheroidals of the Local Group, no similar
work has been done in order to measure biases caused by non-sphericity in this approach. Until now only
\citet{jardel_2012} attempted to take into account the ellipticity when modelling the Fornax dSph. However, in this case
they assumed a particular orientation of the galaxy, avoiding an additional parameter.

As we have demonstrated in \citet{kowalczyk_2017}, the Schwarzschild method recovers the mass and anisotropy profiles
reasonably well for spherical objects. In order to extend the applicability of the method, we decided to test our
procedure on simulated data for a dwarf spheroidal galaxy, measuring the influence of the line of sight: on the
recovered anisotropy under the assumption of the known density profile and on the recovery of both mass and anisotropy
profiles. It will enable us to better understand and assess the validity of future results on modelling observational
data.

The paper is organized as follows. In Sec.\,\ref{merger} we introduce the numerical simulation and
describe the properties of the galaxy used in this study to generate the mock observations in Sec.\,\ref{data}. In
Sec.\,\ref{large_sample} we present our modelling method and apply it to the mock data, and in Sec.\,\ref{small_sample}
we limit the data samples to the size that is currently available observationally. We summarize and discuss the results
in Sec.\,\ref{summary}.

\section{Major merger simulation}
\label{merger}

\citet{kazantzidis_2011b}, \citet{lokas_2014} and \citet{ebrova_2015} have shown that major mergers of two disky
dwarfs can produce realistic dwarf spheroidal galaxies. Therefore, for the purpose of this study we simulated a
collision of two identical dwarfs, each initially composed of an exponential stellar disc with the total mass
$M_s=2\times10^7$\,M$_{\sun}$, the scale-length $R_s=0.41$\,kpc and thickness $z_s/R_s=0.2$ embedded within a
Navarro-Frenk-White-like (NFW, \citealt{NFW_1997}) spherical dark matter halo of virial mass
$M_{DM}=10^9$\,M$_{\sun}$ and concentration $c=20$. The $N$-body realizations were generated using procedures of
\citet{widrow_2005} and \citet{widrow_2008}. Each component of each dwarf was built of $2\times 10^5$ particles, giving
$8\times 10^5$ particles in total. We ran the simulation using $N$-body code GADGET-2 \citep{springel_2005} for the
total time of 10\,Gyr, saving 201 outputs. The adopted softening scales were $\epsilon _s=0.02$\,kpc and $\epsilon
_{DM}=0.06$\,kpc for stellar and dark matter particles, respectively.

At the beginning of the simulation the dwarfs were placed at the relative distance of $d=50$\,kpc and had the relative
radial velocity $v_{orb}=16$\,km\,s$^{-1}$. By assigning no tangential component to the relative velocity we skipped
the inspiralling phase of merging. The angular momentum vectors of the discs were inclined by 45 deg with respect to
the plane of collision and by 90 deg with respect to each other. The galaxies merged during their 3rd approach, after
$t_{merge}=3.7$\,Gyr from the beginning of the simulation.

\begin{figure*}
\includegraphics[trim=15 10 15 25, clip, width=\textwidth]{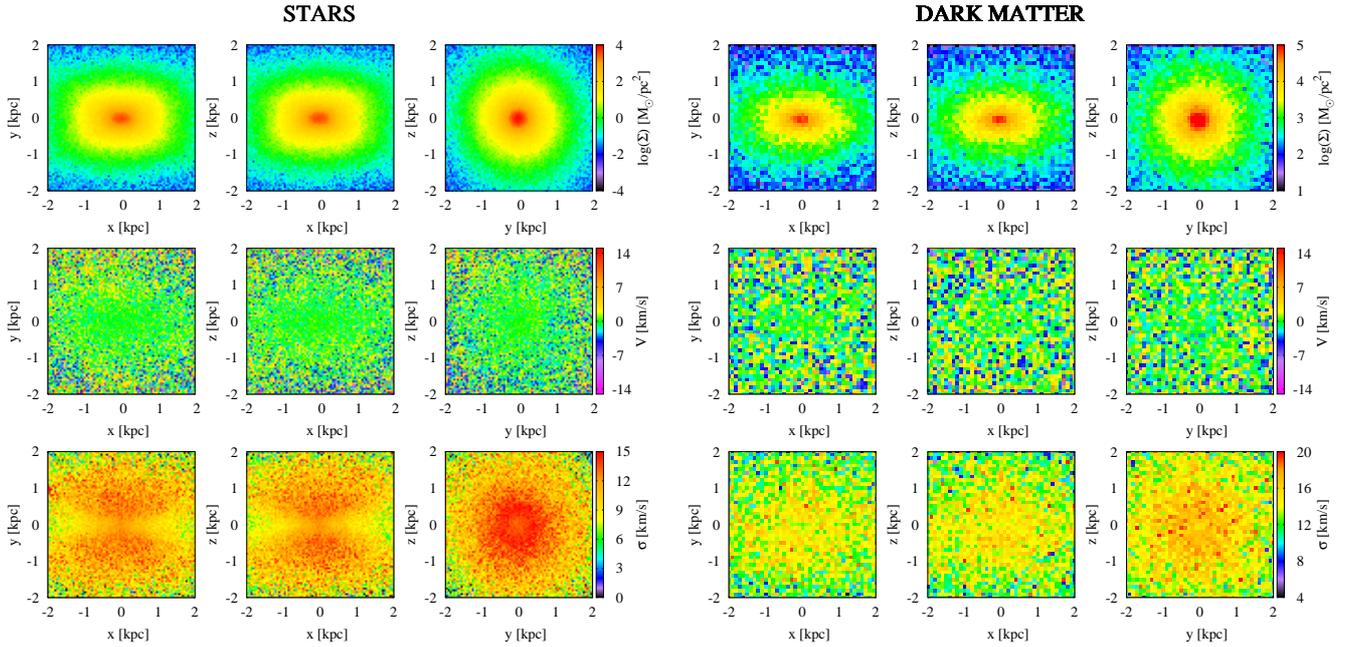}
\caption{Colour maps of the observed parameters of stars (3 $\times$ 3 panels on the left-hand side) and dark matter
(right-hand side) in the final output of the simulation of the major merger. The resolution for dark matter is lower as
the number of particles in the central part of the galaxy is smaller. {\it In rows:} projected
mass density (in logarithm), line-of-sight velocity and line-of-sight velocity dispersion, respectively. {\it In
columns:} observations along the shortest axis $z$, intermediate $y$ and longest $x$.}
\label{fig:maps_all}
\end{figure*}

For further analysis we used the last snapshot from the simulation taken after 10\,Gyr from the beginning when the
galaxy is well relaxed. In Fig.\,\ref{fig:maps_all} we present the colour maps of the surface mass density,
line-of-sight velocity and line-of-sight velocity dispersion (from top to bottom) for the observations along the three
principal axes of the stellar component: the shortest $z$, intermediate $y$ and longest $x$ (in columns from the left to
the right) for stars and dark matter particles (left and right-hand side panels, respectively). As the number of dark
matter particles in the central part of the galaxy ($4 \times 4$\,kpc) is smaller, we reduced the resolution of
these maps by a factor of 4.

Both the stellar component and the dark matter halo are elongated in a similar way with only a few degrees offset between
the directions of principal axes. The axis ratios measured within the radius of 0.5\,kpc are: shortest to longest
$c/a=$0.84 (0.83 for dark matter) and shortest to intermediate $c/b=$0.98 (0.99 or 1.01 for dark matter, as the axes are
switched with respect to the stellar component).

In contrast to \citet{lokas_2014} and \citet{ebrova_2015} who aimed to obtain prolate rotation (rotation around
the major axis of the remnant) as a result of the merger, our galaxy retained no net rotation as the components of the
angular momentum vectors of the dwarfs along the axis of collision had opposite directions. Despite the small
ellipticity of the remnant, the line-of-sight velocity dispersion map reveals interesting, butterfly-like shape for the
lines of sight towards the galaxy perpendicular to the major axis, probably due to the long-axis tube orbits characteristic
of prolate spheroids. It may strongly affect the results of any
spherically symmetric modelling with respect to axisymmetric models. However, any attempt to account for this is beyond
the scope of the present paper.

\section{Mock data}
\label{data}

By observing the galaxy along each of the principal axes of the stellar component we obtained three mock datasets.
For each direction we saved the projected distances from the centre of the galaxy and the line-of-sight velocities of
the stars. We will refer to the datasets for observations along the longest/intermediate/shortest axis as `$x/y/z$
axis'.

\begin{figure}
\begin{center}
\includegraphics[trim=10 35 15 20, clip, width=\columnwidth]{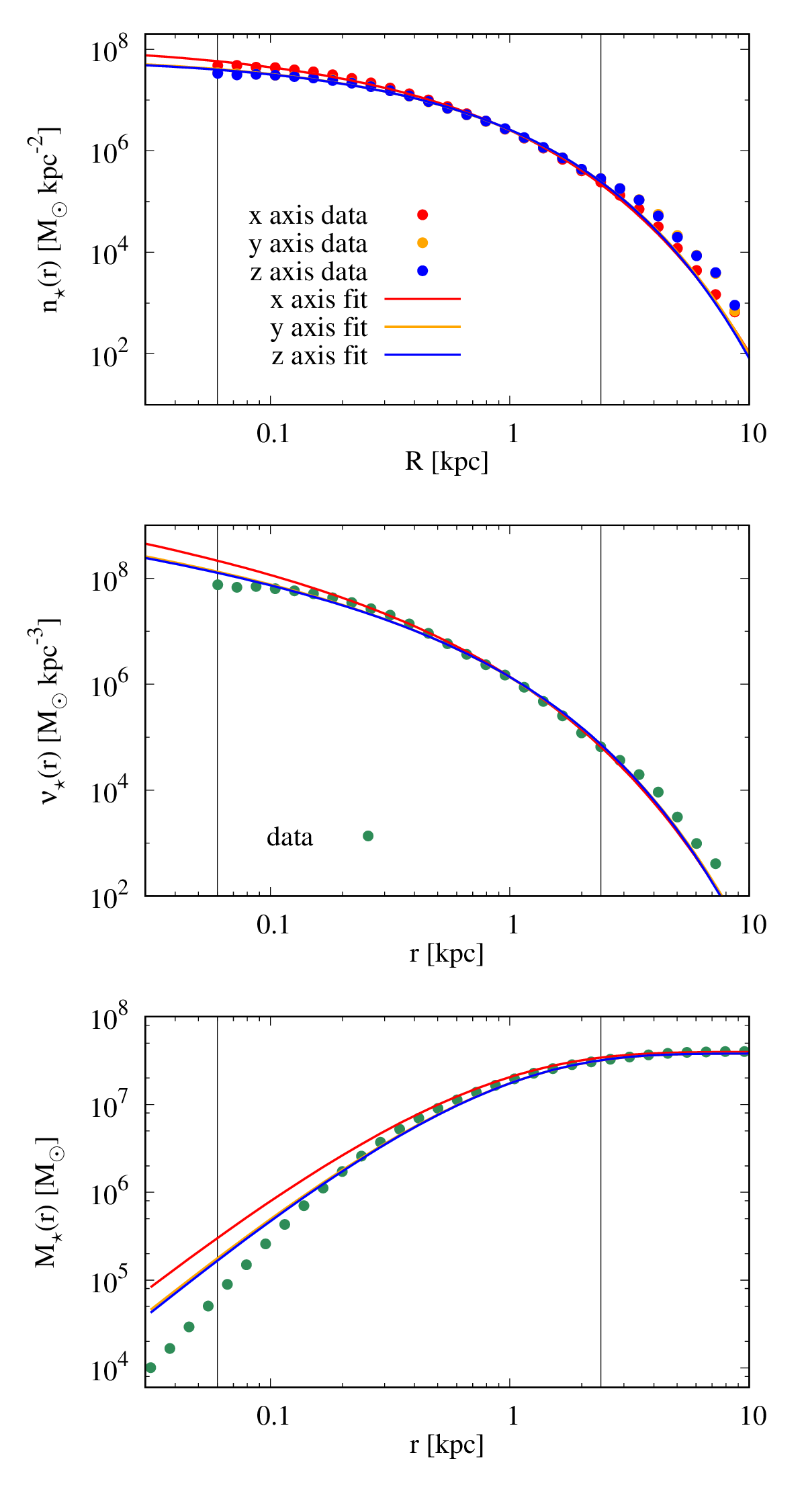}
\caption{{\it Top panel:} the surface mass density of stars as a function of projected radius from the centre of the
galaxy. Points indicate the measurements from the simulation whereas solid lines present the best-fitting
S\'ersic profiles. Colours denote the observations along different principal axes of the stellar component of
the galaxy: red, orange and blue for the longest, intermediate and shortest axis, respectively. {\it Middle
and bottom panels:} the density and cumulative mass of stars as a function of radius from the centre
of the galaxy. Green dots indicate the measurements from the simulation whereas the solid lines present the
analytical formulae with the parameters of the best-fitting profiles. Vertical lines mark the
range in which the profiles were fitted.}
\label{fig:niu_M}
\end{center}
\end{figure}

We present the surface mass density profiles of stars with points in the top panel of Fig.\,\ref{fig:niu_M}, where
colours: red, orange and blue, denote the line of sight: $x$, $y$ and $z$, respectively. We modelled them
with the S\'ersic distribution \citep{sersic_1968}:
\begin{equation}
 n_{\star}(R)=n_0\,{\rm exp}[-(R/R_s)^{1/m}],
\end{equation}
where $n_0$ is the normalization, $R_s$ is the characteristic radius and $m$ is the S\'ersic index. The best-fitting
profiles for each line of sight are shown in Fig.\,\ref{fig:niu_M} with lines of corresponding colour and their
parameters are listed in Table\,\ref{tab:param}. Thin vertical lines restrict the radial range of fitting: the inner
boundary corresponds to minimal credible spatial separation, i.e. 3 softening lengths for the stellar particles in the
simulation whereas the outer one mimics the limiting surface brightness in observations and cuts off visible tails.

\begin{table}
\caption{Parameters of the best-fitting S\'ersic profiles.}
\label{tab:param}
\begin{tabular}{lccc}
\hline
parameter & $x$ axis & $y$ axis & $z$ axis\\
\hline
\multicolumn{4}{c}{large samples (all stars)}\\
\hline
normalization, $n_0$ [$10^7\,$M$_{\sun}$\,kpc$^{-2}$] & 13.60 & 7.40 & 7.08 \\
characteristic radius, $R_s$ [kpc] & 0.080 & 0.132 & 0.145 \\
S\'ersic index, $m$ & 1.825 & 1.661 & 1.620 \\
total mass, $M_s$ [$10^7\,$M$_{\sun}$] & 3.973 & 3.800 & 3.803 \\
\hline
\multicolumn{4}{c}{small samples (100\,000 stars)}\\
\hline
normalization, $n_0$ [$10^7\,$M$_{\sun}$\,kpc$^{-2}$] & 13.49 & -- & 6.95 \\
characteristic radius, $R_s$ [kpc] & 0.082 & -- & 0.148 \\
S\'ersic index, $m$ & 1.818 & -- & 1.607 \\
total mass, $M_s$ [$10^7\,$M$_{\sun}$] & 3.973 & -- & 3.795 \\
\hline
\end{tabular}
\end{table}

Under the assumption of spherical symmetry, any S\'ersic distribution can be deprojected \citep{lima_1999}.
The 3-dimensional density profile is then given as:
\begin{equation}
\label{eq:niu_star}
\nu_{\star}(r)=\nu_0\Big(\frac{r}{R_s}\Big)^{-p}{\rm exp}\bigg[-\Big(\frac{r}{R_s}\Big)^{1/m}\bigg],
\end{equation}
where
\begin{equation}
 \nu_0=\frac{n_0\Gamma(2m)}{2R_s\Gamma\Big((3-p)m\Big)},
\end{equation}
\begin{equation}
 p=1-0.6097/m+0.05463/m^2,
\end{equation}
and $\Gamma(x)$ is the standard gamma function. We present the density profiles resulting from our surface density
fits in the middle panel of Fig.\,\ref{fig:niu_M}. In addition we show the real 3-dimensional stellar mass density
measured from the simulation with green points.

By integrating eq.\,(\ref{eq:niu_star}) over a spherical volume we obtain the profile of the cumulative mass of stars:
\begin{equation}
 M_{\star}(r)=M_s\, \gamma \bigg((3-p)m,\, \Big(\frac{r}{R_s}\Big)^{1/m}\bigg),
\end{equation}
where
\begin{equation}
 M_s=\int\limits_{0}\limits^{\infty} 4\pi r^2 \nu_{\star}(r)\, \mathrm{d}r
\end{equation}
is the total mass of stars and $\gamma(\alpha,\,x)$ is the normalized incomplete gamma function defined as:
\begin{equation}
 \gamma(\alpha,\,x)=\frac{1}{\Gamma(\alpha)}\int\limits_{0}\limits^{x} e^{-t}t^{\alpha-1}\, \mathrm{d}t .
\end{equation}
The profiles of the cumulative mass of stars for our best-fitting S\'ersic profiles are shown with lines of
different colours in the bottom panel of Fig.\,\ref{fig:niu_M} together with the measurements from the
simulation in green. The derived total masses are also given in Tab.\,\ref{tab:param}.

\begin{figure}
\begin{center}
\includegraphics[trim=10 5 30 35, clip, width=\columnwidth]{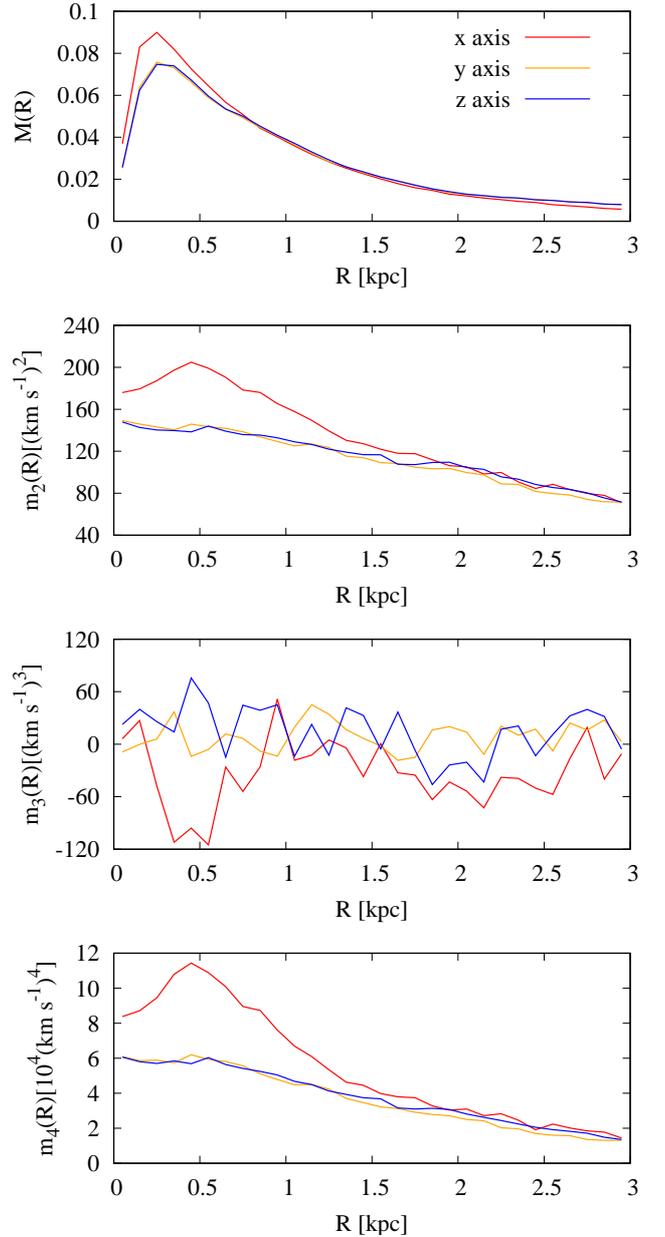}
\caption{The fraction of the stellar mass projected along the line of sight and the 2nd, 3rd, and 4th velocity
moments (top to bottom panels, respectively) derived from all stellar particles for the observations along
the longest (red), intermediate (orange) and shortest (blue) principal axis.}
\label{fig:obs_30}
\end{center}
\end{figure}

We express the kinematics of a dataset in terms of proper moments of the line-of-sight velocity: the second
($m_2$), third ($m_3$) and fourth ($m_4$), calculated with estimators based on the sample of $N$ line-of-sight
velocity measurements $v_i$ \citep{lokas_2003}:
\begin{equation}
 m_{n, l}=\frac{1}{N_l}\sum_{i=0}^{N_l}(v_{i}^{l}-\bar{v_l})^n,
 \label{eq:mom}
\end{equation}
where
\begin{equation}
 \bar{v_l}=\frac{1}{N_l}\sum_{i=0}^{N_l}v_{i}^{l}
 \label{eq:mean}
\end{equation}
and $l$ labels the radial bins. We present the profiles of the moments derived in 30 radial bins from all
stellar particles in Fig.\,\ref{fig:obs_30} in the second, third and fourth panel. Colours denote the line of sight.
The top panel shows the fraction of stars in a given bin, which will be also needed for the modelling.

As both the density profiles and kinematics of the datasets for the observations along the intermediate and
shortest axis are almost identical, we decided to study only two lines of sight: along the longest axis $x$ and
the shortest $z$.

\section{Modelling large datasets}
\label{large_sample}

In this section we shortly describe our modelling method and its application to the mock data obtained by observing a
spheroidal remnant of a major merger simulation along different lines of sight. Our datasets are presented in
Sec.\,\ref{data}.
We examined the accuracy of the recovered mass and orbit anisotropy profiles for observations along the longest and
shortest principal axis in order to determine the bias caused by the non-sphericity of the tracer.

We adopted the definition of the orbit anisotropy of stars given with the anisotropy parameter \citep{GD}:
\begin{equation}
\label{beta}
 \beta(r)=1-\frac{\sigma_{\theta}^2(r)+\sigma_{\phi}^2(r)}{2\sigma_r^2(r)}
\end{equation}
where $\sigma_{r,\,\theta,\,\phi}$ are the components of the velocity dispersion in the spherical coordinate system
with the origin at the centre of the galaxy. In spite of the axisymmetric shape of the galaxy, we have confirmed that
the measured profile does not depend on the orientation of the adopted spherical coordinate system. Throughout the
paper we will compare the results of the modelling to this real anisotropy.

\subsection{Methodology}
\label{method}

In this study we model the ratio of the total density distribution (of stars and dark matter) to the stellar density
with the mass-to-light ratio varying with radius from the centre of a galaxy:
\begin{equation}
 \Upsilon (r)=\frac{\nu_{tot}(r)}{{\nu}_{\star}(r)},
\end{equation}
where $\nu_{tot}(r)$ is the total mass density and ${\nu}_{\star}(r)$ is the
stellar mass density. As both quantities are given in M$_{\sun}$\,kpc$^{-3}$, $\Upsilon(r)$ is dimensionless. However,
assigning a typical mass-to-light ratio of 1 M$_{\sun}$/L$_{\sun}$ to the stellar component, $\Upsilon (r)$ can be
expressed in solar units, as is typically done in comparisons with observations.

\begin{figure}
\begin{center}
\includegraphics[trim=10 10 20 40, clip, width=\columnwidth]{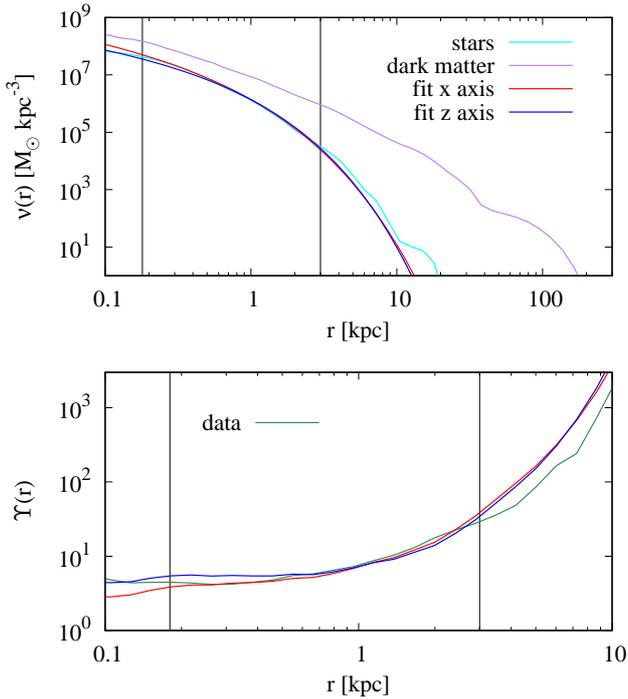}
\caption{{\it Top panel:} the comparison of density profiles of stars (cyan line) and dark matter (purple line).
Additionally, in red and blue we show the deprojected S\'ersic profiles derived for the parameters of the best-fitting
surface mass density profiles for observations along the longest and shortest axis. Black vertical lines mark
the 3 softening scales for the dark matter
and the outer radius of the data sets (from left to right, respectively). {\it Bottom panel:} the ratio between
the total mass density (stars and dark matter) and stellar mass density as a function of radius. The green
line presents the true values whereas the red and blue lines indicate the same ratio under the assumption that the
stellar densities are given with the best-fitting S\'ersic profiles.}
\label{fig:m_l}
\end{center}
\end{figure}

We present the comparison of density profiles of stars and dark matter in the top panel of Fig.\,\ref{fig:m_l} with
cyan and purple lines, respectively. The dark matter halo is much more extended and up to $\sim 20$\,kpc can be
well approximated with an NFW profile with the virial mass $M_{v, rem}=1.39\times 10^9$\,M$_{\sun}$ and the
concentration $c_{rem}=21.8$. We also reproduce here the best-fitting S\'ersic profiles for the two studied lines of
sight in red and blue as in the bottom panel we show the profiles of mass-to-light ratio under an assumption that the
stellar density follows the deprojected S\'ersic profiles. As they depend on the parameters of a S\'ersic profile,
they depend also on the line of sight. We compare the profiles with the real mass-to-light ratio from the
simulation, shown in green, which is additionally affected by a tail in the stellar distribution.

We decided to use the deprojection instead of the fit to the 3-dimensional density profile in order to avoid
introducing additional parameters when the mass-to-light ratio is not known a priori. As shown in Fig.\,\ref{fig:m_l}
the deprojections reproduce the real profile well in the considered radial range (between the vertical lines), but in
the centre of the galaxy the density is overestimated. However, a core in the stellar distribution may not be of any
physical nature as it appears within a radius corresponding to 3 softening scales for the dark matter particles,
rather suggesting a numerical artifact.

We modelled our data by applying the spherically symmetric Schwarzschild orbit superposition method
\citep{schwarzschild_1979}. We described and justified the details of our approach and tested its reliability in
recovering the anisotropy and total mass profiles for spherical objects in \citet{kowalczyk_2017}. Therefore, in this
work we summarize only the crucial steps of the method:
\begin{enumerate}
 \item For a given total density profile we generate a set of initial conditions for an orbit library.
 In order to properly sample orbits available in this potential, a library needs to be representative in energy and angular
 momentum spaces. We use 100 values of energy in units of the radius of the circular orbit sampled logarithmically and
 12 values of the relative angular momentum $l=L/L_{\rm max}$, where $L_{\rm max}$ is the angular momentum of the circular
 orbit, sampled linearly within the open interval $l\in(0, 1)$ to avoid numerical errors. Additionally, a library needs to
 spatially cover over 99.9\% of the mass of a tracer (based on a fitted S\'ersic profile) which puts a lower limit on the
 outer radius of a library and therefore the maximum value of energy. The minimum value is chosen so that the apocentres of
 corresponding orbits are smaller than the outer radius of the innermost bin used in modelling. We integrate orbits using
 $N$-body code GADGET-2 \citep{springel_2005} saving 2001 outputs in
 equal timesteps with the total integration time adjusted to cover at least few orbital periods.\label{en:step_1}

 \item We randomly rotate each coplanar orbit 100\,000 times around two axes of the simulation box, combining them to
 mimic the spherical symmetry and observe each orbit along an arbitrarily chosen line of sight saving the same observables
 in the same radial binning as for the data (see Sec.\,\ref{data}).
 Additionally, we store three components of velocity dispersion
 in the spherical coordinate system and deprojected fractions of mass, i.e. fractions in 3-dimensional shells. For orbits
 we identify the fraction of mass (projected and deprojected) in a given bin as the fraction of the total integration
 time spent in the bin.

 \item We fit the observables of an orbit library to the data by assigning non-negative weights $\gamma$ to orbits so that
 their linear combination minimizes the objective function $\chi^2$:
 \begin{equation}
\label{eq:fit}
 \chi^2=\sum_{l}\sum_{n}\Bigg(\frac{M_l^{\rm obs}m_{n,l}^{\rm obs}-\sum_k\gamma_kM_l^km_{n,l}^{k}}{\Delta
(M_l^{\rm obs}m_{n,l}^{\rm obs})}\Bigg)^2
\end{equation}
under the constraints that for each orbit $k$ and each bin $l$:
\begin{equation}
\label{eq:weights}
\left\{
\begin{array}{l}
|M_l^{\rm obs}-\sum_k\gamma_kM_l^k|\leq\Delta M_l^{\rm obs}\\
\gamma_k\ge 0\\
\sum_k\gamma_k=1
\end{array} \right.
\end{equation}
where $M_l^k$, $M_l^{\rm obs}$ are the fractions of the projected mass of the tracer contained within $l$th bin for
$k$th orbit or from the observations and $m_{n,l}^k$, $m_{n,l}^{\rm obs}$ are $n$th proper moments. $\Delta$ denotes
the measurement uncertainty associated with a given parameter. The velocity moments are weighted with the projected
masses and to derive the errors we treat both quantities as independent.
We execute the $\chi^2$ fitting of eq.\,(\ref{eq:fit}) with rigid constraints of eq.\,(\ref{eq:weights}) with the
non-negative quadratic programming (QP) implemented in the CGAL library \citep{cgal}. \label{en:step_3}

\item We derive the anisotropy resulting from the modelling assuming that the velocity dispersions in the spherical
coordinate system are also given with the linear combinations of orbital dispersions weighted with the 3-dimensional mass
fractions. Therefore the anisotropy in $l$th bin is:
\begin{equation}
 \beta_l=1-\frac{\sum_k\gamma_kM_{{3D}, l}^k(\sigma_{\theta, l}^k)^2+\sum_k\gamma_kM_{{3D}, l}^k(\sigma_{\phi,
l}^k)^2}{2\sum_k\gamma_kM_{{3D}, l}^k(\sigma_{r, l}^k)^2}
\end{equation}
where $\sigma^k_{(r,\,\theta,\,\phi),l}$ are the components of the velocity dispersion for the $k$th orbit calculated
in the $l$th spatial bin. We consider only $l>1$ as the result in the innermost bin cannot be trusted.

\item The best-fitting total density profile is determined as the minimum among the absolute values of $\chi^2$
function on a grid of density profiles created by varying the free parameters of the profile and for each of their
combinations repeating steps \ref{en:step_1}-\ref{en:step_3}.

The confidence levels for the recovered density profile with two free parameters (in this work we use $a$ and
$\Upsilon_0$, see Sec.\,\ref{upsilon}) are derived by fitting surfaces of 4th or 8th order to the $\chi^2$ maps ($\sim
a^2\Upsilon_0^2$ or $\sim a^4\Upsilon_0^4$ depending on the level of complexity of the map) and applying standard
$\chi^2$ statistics for two degrees of freedom, i.e. $\Delta\chi^2=\chi^2-\chi_{min}^2=2.30,\,6.17,\,11.80$
corresponding to 1,\,2,\,3\,$\sigma$ \citep{NR}. Fitting of a surface is necessary due to numerical artifacts affecting
the method (see \citealt{kowalczyk_2017} for details).
\end{enumerate}

Throughout the paper we will refer to the values of various parameters obtained from the full 6D data
from the simulation as \textit{real} and those obtained with the modelling of mock data with the
Schwarzschild method as \textit{derived}.

\subsection{Constant mass-to-light ratio}

First, we considered the simplest scenario in which {\it mass-follows-light}, i.e. $\Upsilon(r)=\Upsilon_0=$~const.
It is a good assumption e.g. for strongly tidally stripped galaxies orbiting their hosts on tight orbits \citep{lokas_2013}.

In this case we integrated only one orbit library for each line of sight with the total density profile matching
the deprojected best-fitting S\'ersic profile. Libraries for different values of $\Upsilon_0$, a
constant scaling of mass, were obtained with a simple transformation resulting from the energy conservation and performed
on each orbit separately \citep{rix_1997}:
\begin{eqnarray}
\label{eq:trans}
v_i&\longrightarrow &\sqrt{\Upsilon_0}v_i\\
\bar{v_l}&\longrightarrow &\sqrt{\Upsilon_0}\bar{v_l}\nonumber \\
m_{n,l}&\longrightarrow &\Upsilon_0^{\frac{n}{2}}m_{n,l}\nonumber \\
\beta_l&\longrightarrow &\beta_l\nonumber
\end{eqnarray}
where $v_i$ is a single line-of-sight velocity measurement on an orbit, $\bar{v_l}$ is the mean velocity in $l$th bin
and $m_{n,l}$ is the $n$th proper moment. Since the anisotropy is defined with a ratio of velocity dispersions, the
multiplication factor gets cancelled. Therefore, the transformation of the velocity dispersions in the spherical coordinate
system is not necessary.

\begin{figure}
\begin{center}
\includegraphics[trim=15 0 20 12, clip, width=\columnwidth]{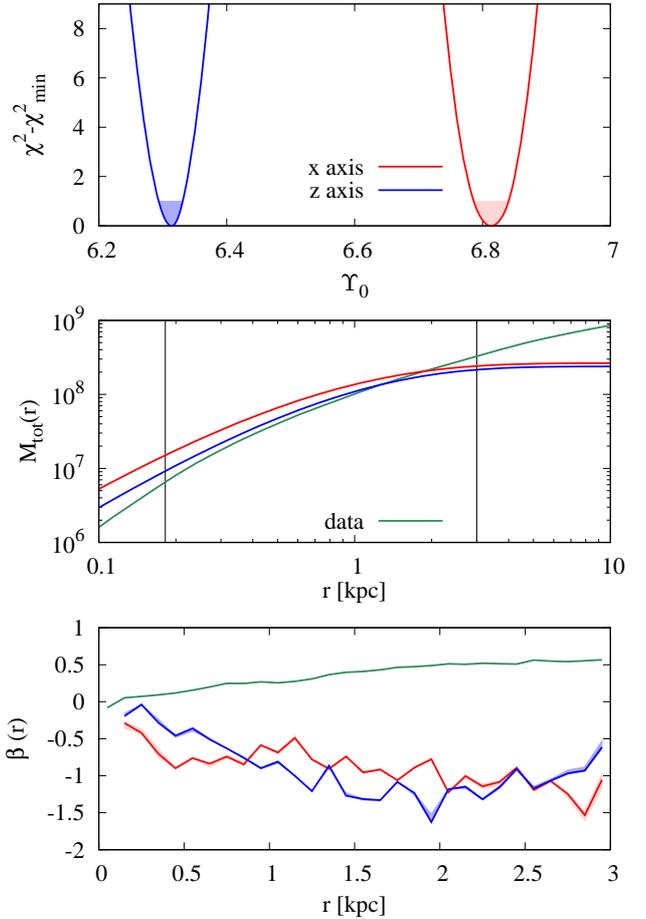}
\caption{The results of the Schwarzschild modelling under the assumption of the mass-to-light ratio being
constant with radius. {\it Top panel:} the distribution of $\Delta \chi^2$ as a function of mass-to-light
ratio for the data based on observations along the longest (red) and shortest (blue) axis. The minima of the
distributions correspond to the best-fitting models whereas the shaded ranges of the corresponding colour
show 1\,$\sigma$ uncertainties. {\it Middle panel:} the total mass profiles from the simulation (in green)
and for the best-fitting models (red and blue lines) together with the 1\,$\sigma$ error bars (shaded regions,
indistinguishable in the scale of the plot). Black vertical lines mark the 3 softening scales
for the dark matter and the outer radius of the data sets (from left to right, respectively). {\it Bottom
panel:} the anisotropy parameter profiles. The shaded regions correspond to the extreme values for the
mass-to-light ratios contained within 1\,$\sigma$ confidence level.}
\label{fig:Upsilon_const}
\end{center}
\end{figure}

We present the results of our modelling in Fig.\,\ref{fig:Upsilon_const}. The top panel shows the
$\Delta\chi^2=\chi^2-\chi^2_{min}$ curves as a function of mass-to-light ratio for the observations along the longest
(in red) and shortest (in blue) principal
axis of the stellar component. Shaded areas identify the 1\,$\sigma$ confidence levels $\Delta\chi^2\leq 1$ for
1 degree of freedom \citep{NR}. The resulting total mass profiles are shown in the middle panel together with the
real mass profile derived from the simulation in green. The differences in the profiles between the best-fitting
models (corresponding to the minimum of each curve) and the models contained within 1\,$\sigma$ are smaller than
the thickness of lines in the figure. Thin vertical lines indicate 3 softening scales for dark matter particles and
the outer radius of data sets from left to right, respectively.

The bottom panel of Fig.\,\ref{fig:Upsilon_const} presents the anisotropy profiles: resulting from the modelling and
calculated from the full data from the simulation for comparison. Colours are consistent between the
panels.

Since our data clearly cannot be well approximated with a mass-follows-light model, such an assumption leads to the
underestimation of mass content at large radii and the derived anisotropy strongly biased towards
tangential orbits. For both lines of sight the anisotropy profiles decrease with radius and the discrepancies reach
$\Delta\beta\approx 2$ at $r\geq 2$\,kpc.

\subsection{Mass-to-light ratio varying with radius}
\label{upsilon}

We generalized our approach by allowing the mass-to-light ratio to vary with radius, following a formula:
\begin{equation}
\label{eq:Upsilon}
\log\Upsilon(r) = \left\{
\begin{array}{ll}
c & \log r\leq -0.74\\
a(\log r + 0.74)^3 +c & \log r> -0.74\\
\end{array} \right.
\end{equation}
where $\Upsilon(r)$ is dimensionless and $r$ is given in kpc. The parameters $a$ and $c$ are constants defining a
density model.

Our formula represents a cubic curve in the log-log scale with the minimum at the radius corresponding to
approximately three softening scales for the dark matter halo and constant for smaller radii. Since the case
of $a=0$ reduces to the mass-follows-light model studied in
the previous section, we will use $\Upsilon_0=10^c$ and we will refer to $a$ as a `curvature parameter'. We
consider only $a\ge0$, excluding cases  where the density of the dark matter halo drops faster
at large radii than the density of stars, which is not supported by any numerical experiments. Note also that
our definition of the mass-to-light ratio assumes that in the centre the dark matter follows the stellar
density distribution, i.e. it has the same mild cusp as the deprojected S\'ersic profile.

In order to determine the true mass profiles, i.e. the combinations of parameters $a$ and $\Upsilon_0$ reproducing the
real profiles most accurately, and quantify the bias caused only by the non-spherical shape of the galaxy, we fitted
the mass-to-light ratio profiles to the full data from the simulation under the assumption that the stellar density
profiles follow the deprojected best-fitting S\'ersic distribution. The reason we use the deprojected best-fitting S\'ersic
for each line of sight is because that is what is actually observable. The deprojections are different and this
difference propagates into the estimate of $\Upsilon$.

\begin{figure}
\begin{center}
\includegraphics[trim=15 5 30 0, clip, width=\columnwidth]{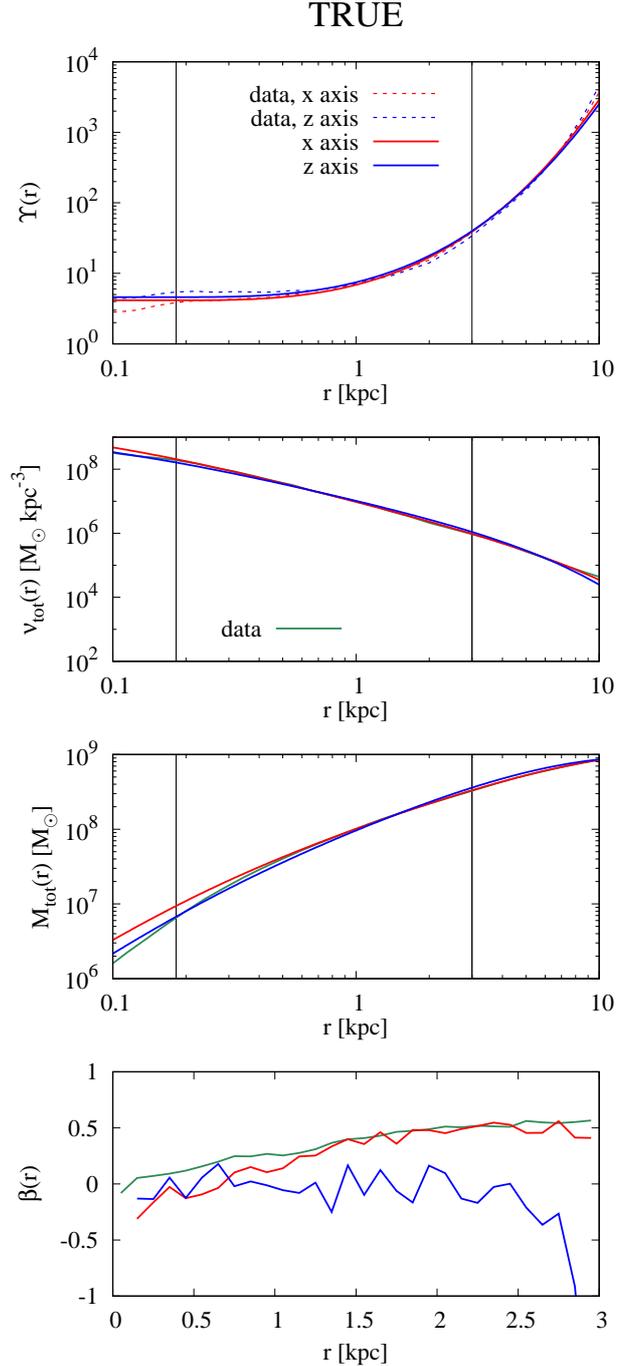}
\caption{Results of Schwarzschild modelling for all particles under the assumption of the best-fitting mass-to-light
ratio models presented in the top panel with solid lines: red for the observations along the longest and blue
for the shortest principal axis. Dashed lines indicate the real profiles from the simulation under the assumption
that the stellar distribution is given with the best-fitting S\'ersic profiles. Black vertical lines mark the 3
softening scales for the dark matter and the outer radius of the data sets (from left to right, respectively).
{\it Second, third and forth panels:} profiles of total density, cumulative total mass and anisotropy, respectively.
Green lines present the real values from the simulation.}
\label{fig:Upsilon_true}
\end{center}
\end{figure}

The best-fitting mass-to-light ratio profiles
(`true profiles' hereafter) given with eq.\,(\ref{eq:Upsilon}) are presented in the top panel of
Fig.\,\ref{fig:Upsilon_true} with solid lines for the observations along the longest (in red) and shortest (in blue)
principal axis. Dashed lines show the real profiles from the simulation for comparison (see Fig.\,\ref{fig:m_l}). Thin
vertical lines indicate 3 softening scales for dark matter particles and the outer radius of data sets from left to
right, respectively.
Middle panels of Fig.\,\ref{fig:Upsilon_true} show the mass density and cumulative mass profiles resulting from the
true mass-to-light ratio profiles and we compare them with the real profiles from the simulation (in green).

In the bottom panel we present the anisotropy profiles resulting from the modelling for the datasets obtained by
observations along the longest and shortest axis in red and blue, respectively, whereas the green curve shows the
real anisotropy profile of the simulated galaxy calculated from the full 6D information. For the observations along the
longest axis anisotropy is underestimated only in the centre of the galaxy, where our mass-to-light ratio model
overestimates cumulative mass, but is well recovered at larger radii.

The situation is much worse for the observations along the shortest axis where the derived anisotropy is approximately
0 up to $r\approx 2$\,kpc and drops rapidly beyond whereas the real anisotropy profile is monotonically growing.
This is an effect of the elongated shape of the galaxy in projection. Since we calculate the density and kinematics in
concentric rings, we average the properties of stars which in axisymmetric modelling would belong to different elliptical
shells. Therefore, we can see that even for the correct mass profile, the bias is significant.

\subsection{Recovering mass-to-light ratio profile}
\label{density_30}

\begin{figure}
\begin{center}
\includegraphics[trim=5 20 15 20, clip, width=\columnwidth]{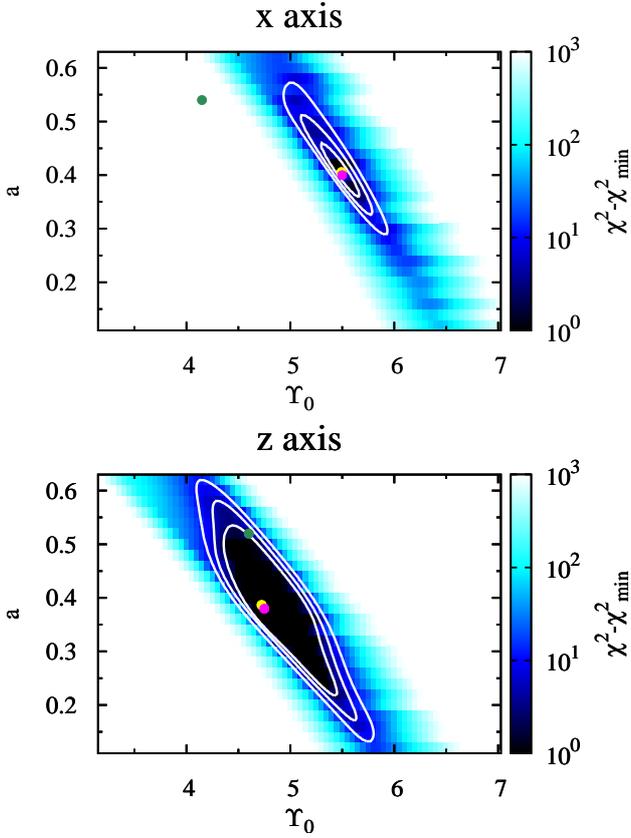}
\caption{The maps of the $\chi^2$ values relative to the minima of fitted surfaces for the data sets obtained
by observing the galaxy along the longest (top panel) and shortest (bottom panel) principal axis on the grids
of different mass-to-light ratio models. The global minima are marked with yellow dots and the true
values with green ones. Magenta points
indicate the best-fitting mass profiles, i.e. the profiles on the grid closest to the global minima along the
contours of equal $\Delta \chi^2$ plotted with white curves.}
\label{fig:chi_30}
\end{center}
\end{figure}

In the next step we determined the reliability of recovering the total density and anisotropy profiles simultaneously. For this
purpose we calculated two sets of orbit libraries (one for each line of sight) by varying the curvature parameter in
the range $a\in[0,\, 0.9]$ with a step $\Delta a=0.02$. Libraries for different values of $\Upsilon_0$ were obtained
using the transformation given with eq.\,(\ref{eq:trans}). We present the colour maps of the absolute values of the
$\chi^2$ function for best-fitting models (relative to the minimum of fitted surface, see Sec.\,\ref{method}) in
Fig.\,\ref{fig:chi_30} where the top panel corresponds to the results for the observations along the longest axis
whereas the bottom panel for the shortest. In both panels the true profile is marked with a green dot while the minimum
of the fitted surface with a yellow one. We identify the best-fitting density model as the profile on the grid closest
to the global minimum along the contours of equal $\Delta \chi^2$. Those models are marked with magenta dots.
Additionally, with white curves we indicate the 1,\,2,\,3\,$\sigma$ confidence levels.  For the observations
along the shortest axis the true profile was recovered within 2\,$\sigma$, whereas for the longest axis the true profile
was not recovered at all (the true profile is far outside 3\,$\sigma$ contour, $\Delta \chi^2$>3\,800). As the interpretation
of bias based on the mass-to-light ratio profile parameters is difficult, we will comment on it when referring to the
total mass profile.

\begin{figure}
\begin{center}
\includegraphics[trim=15 5 30 0, clip, width=\columnwidth]{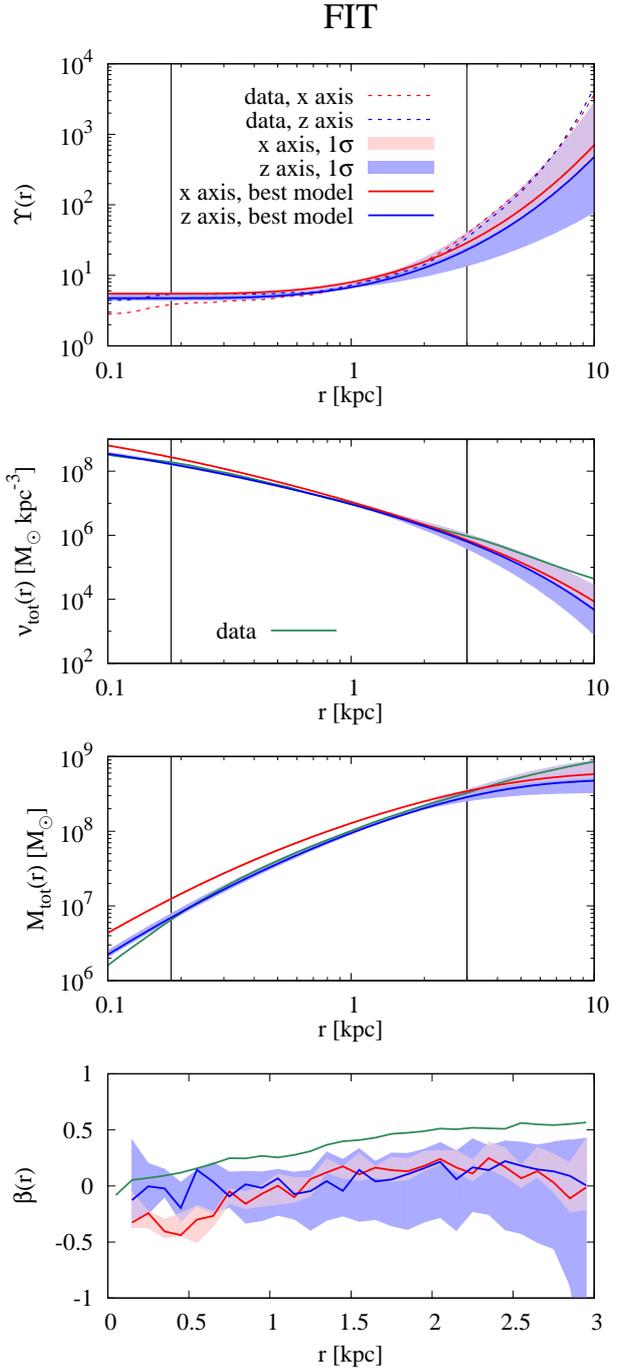}
\caption{Results of Schwarzschild modelling for all stars based on the fitted mass-to-light ratio profiles
{\it First panel:} Mass-to-light ratio profiles for the observations along the longest (red) and
shortest (blue) principal axis. Dashed lines indicate the real profiles from the simulation under the assumption
that the stellar distribution is given by the best-fitting S\'ersic profiles whereas solid lines present the results
for the best-fitting models. Shaded regions denote the spread of values for the models derived within 1\,$\sigma$
confidence level. Black vertical lines mark the 3 softening scales for the dark matter and the outer radius of
the data sets (from left to right, respectively).
{\it Second, third and fourth panel:} profiles of total density, cumulative total mass and anisotropy, respectively.
Green lines present the real values from the simulation.}
\label{fig:Upsilon_fit}
\end{center}
\end{figure}

The profiles of obtained mass-to-light ratio, total density, total mass and anisotropy are presented
in the consecutive
panels of Fig.\,\ref{fig:Upsilon_fit}. Values for the best-fitting models (magenta dots in Fig.\,\ref{fig:chi_30}) are
shown with solid lines: red for the observations along the longest principal axis and blue for the shortest. The ranges
spanned by all the models contained within 1\,$\sigma$ are indicated with light red and light blue shaded areas (or
purple when red and blue areas overlap). In each panel the resulting parameters are compared with the
real values from the simulation represented with dashed lines (mass-to-light ratio profiles only) or green solid lines
(otherwise).

The mass profile is fairly well recovered for the observations along the shortest axis and overestimated for the
longest. At the outskirts of the galaxy ($r>2$\,kpc) the best-fitting models for both lines of sight underestimate
the mass content, however the real profile is enclosed within 1\,$\sigma$.

When the density profile is to be recovered with the method, the bias in the anisotropy occurs for both lines of sight.
For the observations along the longest axis the derived anisotropy profile is growing (for the best-fitting model it has
a local maximum and slightly decreases at larger radii), however the values of anisotropy are systematically
underestimated with the mean offset of $\Delta \beta=0.38$ for the best-fitting model. For the observations
along the shortest axis the best-fitting model is consistent with isotropic orbits ($\bar{\beta}=0.08$) but 1\,$\sigma$
confidence level allows the anisotropy profile to grow (with mean minimal offset with respect to the true values
$\Delta \beta\approx 0.1$) or decrease (as for the true mass-to-light ratio profile).

\section{Small data samples}
\label{small_sample}

In this part of our work we studied the bias in the recovered density and anisotropy profiles caused by the axisymmetric
shape of the galaxy when the data samples used in the modelling are comparable in size with the available data sets for
dwarf galaxies of the Local Group.

\subsection{Examples of data modelling}

We present the results of modelling for two small data samples. They were obtained by observing the remnant of the major
merger simulation along the longest and shortest principal axis of the stellar component (see Sec.\,\ref{data}) and
randomly choosing 100\,000 stars for which only the projected distances from the centre of the galaxy were known and
2\,500 stars with the projected distances and line-of-sight velocities.
This means that the sampling follows the projected light distribution. The data are then binned
into bins of equal size in the projected radius to facilitate the modelling. This results in a different number of
stars per bin and the data are assigned sampling errors according to this number.

\begin{figure}
\begin{center}
\includegraphics[trim=10 5 30 35, clip, width=\columnwidth]{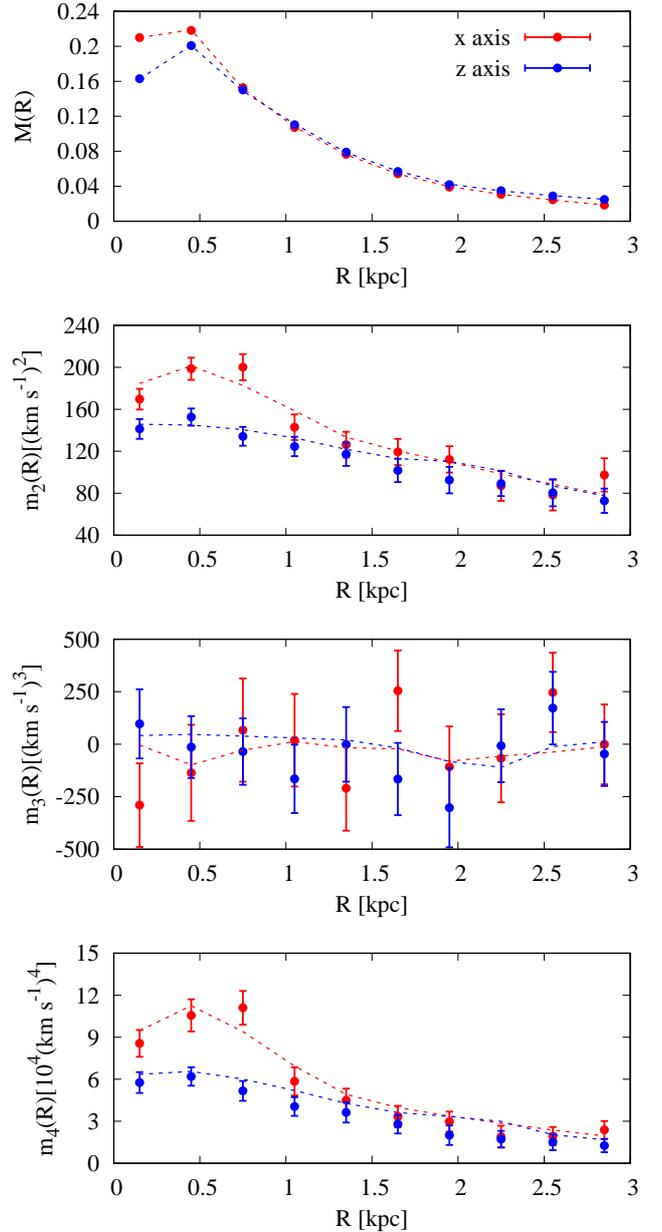}
\caption{The fraction of the stellar mass projected along the line of sight and the 2nd, 3rd, and 4th velocity
moments (top to bottom panels, respectively) for the data sets obtained by observing the galaxy along the
longest (red) and shortest (blue) principal axis. The points with the 1\,$\sigma$ error bars represent the values
for the small random samples with sampling errors, while the thin dashed lines show the measurements based on
all stellar particles from the simulations with the same binning.}
\label{fig:obs_10}
\end{center}
\end{figure}

Fig.\,\ref{fig:obs_10} presents the observables for the selected small samples (points): mass fraction based on
100\,000 stars and velocity moments 2-4 of line-of-sight velocities. The same parameters in the same binning derived
from all stars are shown with dashed lines for comparison. Colours, as before, denote the line of sight: red for the
observations along the longest axis and blue for the shortest. Error bars indicate 1\,$\sigma$ sampling errors which
were determined in Monte Carlo (MC) simulations. For each line of sight and each radial bin we constructed a grid of
sampling errors as a function of the number of particles in a bin. For the mass fraction we assumed Poissonian errors
and they are smaller than data points in Fig.\,\ref{fig:obs_10}.

The parameters of the best-fitting S\'ersic profiles for the chosen subsamples of stars with distances only are quoted
in Table\,\ref{tab:param} and are very similar to the ones derived for all stars. Therefore, we found it unnecessary to
integrate new orbit libraries. In order to guarantee that we measure the same parameters as in
Sec.\,\ref{large_sample}, we assumed that the total luminosity of the galaxy and the stellar mass-to-light ratio
(constant with radius) are known (\citealt{mateo_1998}, \citealt{mcconnachie_2012}).

First, we determined the sampling errors of the derived anisotropy, i.e. the uncertainties in the
derived anisotropy caused by using different small samples, for the true mass-to-light ratio profiles.
For that purpose for each line of sight we applied our Schwarzschild method, i.e. fitted the
orbit library, to 10\,000 random samples and investigated the statistics of anisotropy profiles.
The values of the mean and 1\,$\sigma$ deviation in each bin were obtained by fitting a Gaussian profile to
the histogram of the derived anisotropy.

\begin{figure}
\begin{center}
\includegraphics[trim=20 5 25 10, clip, width=\columnwidth]{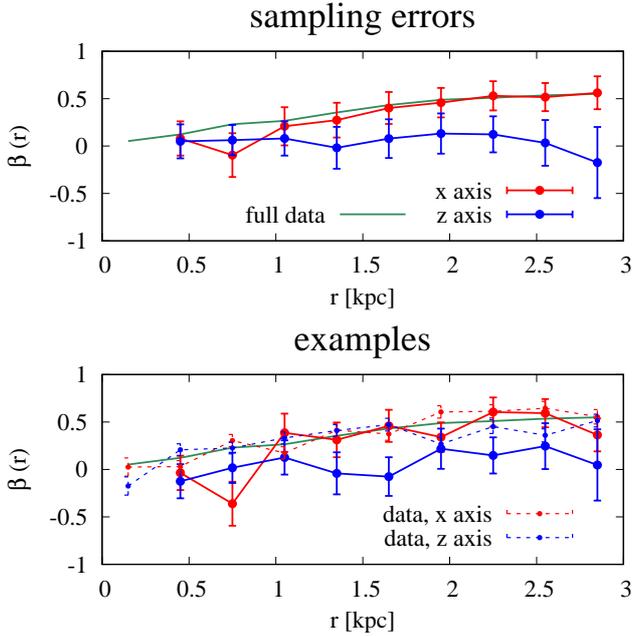}
\caption{{\it Top panel:} Data points present the mean values of the derived anisotropy resulting from
the MC simulations for the true mass-to-light ratio profiles. The red colour denotes the observations along
the longest and the blue along the shortest one. The 1\,$\sigma$ sampling errors are shown with error bars. The real
anisotropy profile derived from the simulation is presented with a green line. {\it Bottom panel:} Examples of modelling
for two small data samples (see text), one for each line of sight under the assumption of the true mass-to-light ratio
profiles. Large data points connected with solid lines indicate the derived values whereas the error bars
represent the 1\,$\sigma$ sampling errors and are the same as in top panel. With small data points and dashed lines
of the corresponding colour we show the real anisotropies calculated using only stars in small
samples. For the real anisotropy the sampling errors were obtained with the MC method.}
\label{fig:beta_true_ss}
\end{center}
\end{figure}

We present
the results of this experiment in the top panel of Fig.\,\ref{fig:beta_true_ss} where the data points show mean values
of the derived anisotropy and error bars denote 1\,$\sigma$ errors. Colours indicate the line of sight: red for the
observations along the longest axis and blue for the shortest. In green we reproduce the real profile of the anisotropy
for comparison.

The mean values of the derived anisotropy show similar trends to those we obtained for the large samples.
For observations along the longest axis the anisotropy is underestimated in the centre and well recovered at
further radii. For observations along the shortest axis the anisotropy profile is consistent with an isotropic
model and decreases slightly at the outskirts of the galaxy (compare with bottom panel in
Fig.\,\ref{fig:Upsilon_true}). Sampling errors for the derived anisotropy averaged over bins are 0.18 and 0.22 for
the $x$ and $z$ axis, respectively.

In the bottom panel of Fig.\,\ref{fig:beta_true_ss} we present the anisotropy profiles (large points and solid lines)
derived with our method for two random small samples, one for each line of sight, for which observables are shown in
Fig.\,\ref{fig:obs_10}. We consistently use red colour for the observations along the longest axis, blue for the
shortest and green for the real values (the latter calculated from all available stellar particles).
The error bars are the same as in the top panel. For comparison of sampling errors between the derived and real
anisotropies, values of the real anisotropy calculated from the small samples and corresponding sampling errors are
shown with small points and dashed lines. As for the velocity moments, sampling errors of the real anisotropy were
determined with MC simulations.

\subsection{Recovering the mass profile}

In the last part of our study we investigated the reliability of recovering both the mass-to-light ratio and
anisotropy profiles for the two small samples described in the previous section. We used the same orbit libraries
and procedures as in Sec.\,\ref{density_30}.

\begin{figure}
\begin{center}
\includegraphics[trim=5 10 15 20, clip, width=\columnwidth]{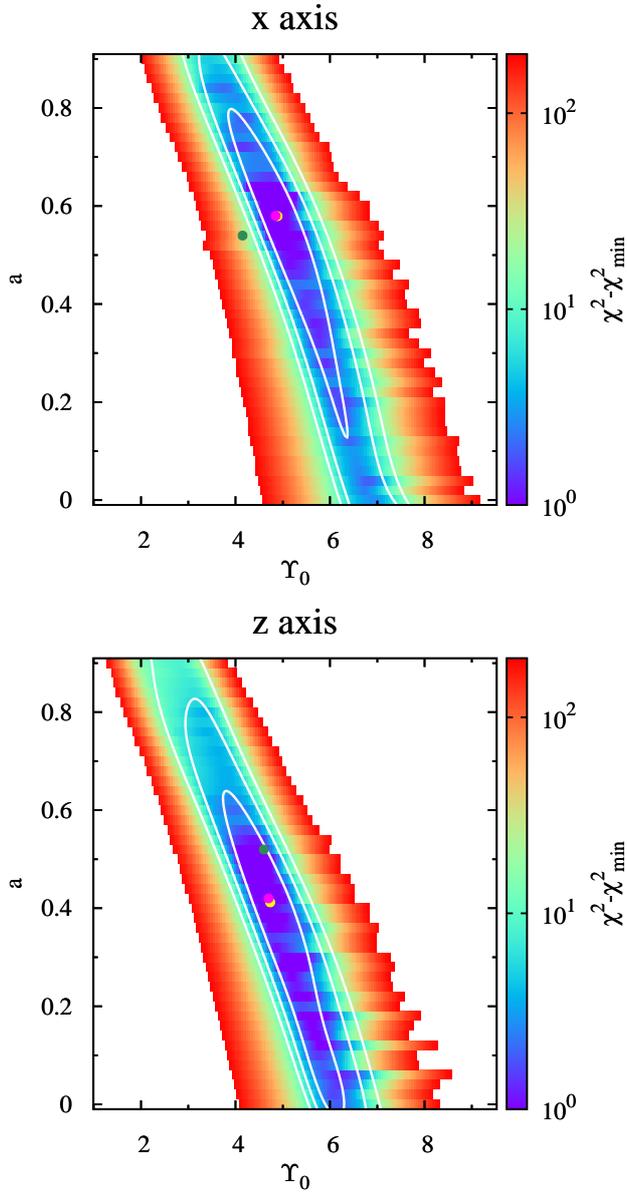}
\caption{Maps of the $\chi^2$ values relative to the minima of the fitted planes for the small data samples
of 100\,000/2\,500 stars. The two panels correspond to the data sets obtained by observing the galaxy along
the longest (top) and shortest (bottom) axis. The global minima are marked with yellow dots whereas the true
values of the mass-to-light ratios with green ones. Magenta dots indicate the best-fitting models identified as
the closest to minimum on the grid.
White lines show the contours of equal $\Delta \chi^2$ corresponding to 1, 2, 3$\sigma$ confidence
levels.}
\label{fig:chi_10}
\end{center}
\end{figure}

Fig.\,\ref{fig:chi_10} presents the colour maps of $\Delta \chi^2$, i.e. absolute values of the objective
function relative to the minimum of a fitted two dimensional surface, as a function of the two parameters of
the mass-to-light ratio profile: the normalization $\Upsilon_0$ and the curvature parameter $a$. The two panels
correspond to the different lines of sight. White loops indicate the 1,\,2,\,3\,$\sigma$ confidence levels based on the
fitted surfaces and yellow dots mark their minima. We identify the best-fitting model as the closest to the minimum on
the grid of mass-to-light ratio profiles and mark them with magenta dots. Green dots show the true profiles, which were
used in the previous section.  Similarly to the large samples, for the observations along the shortest
axis the true mass-to-light ratio profile was recovered within 1\,$\sigma$ (2\,$\sigma$ for large sample) and not
recovered for the longest ($\Delta \chi^2\approx 38$).

\begin{figure}
\begin{center}
\includegraphics[trim=15 5 30 0, clip, width=\columnwidth]{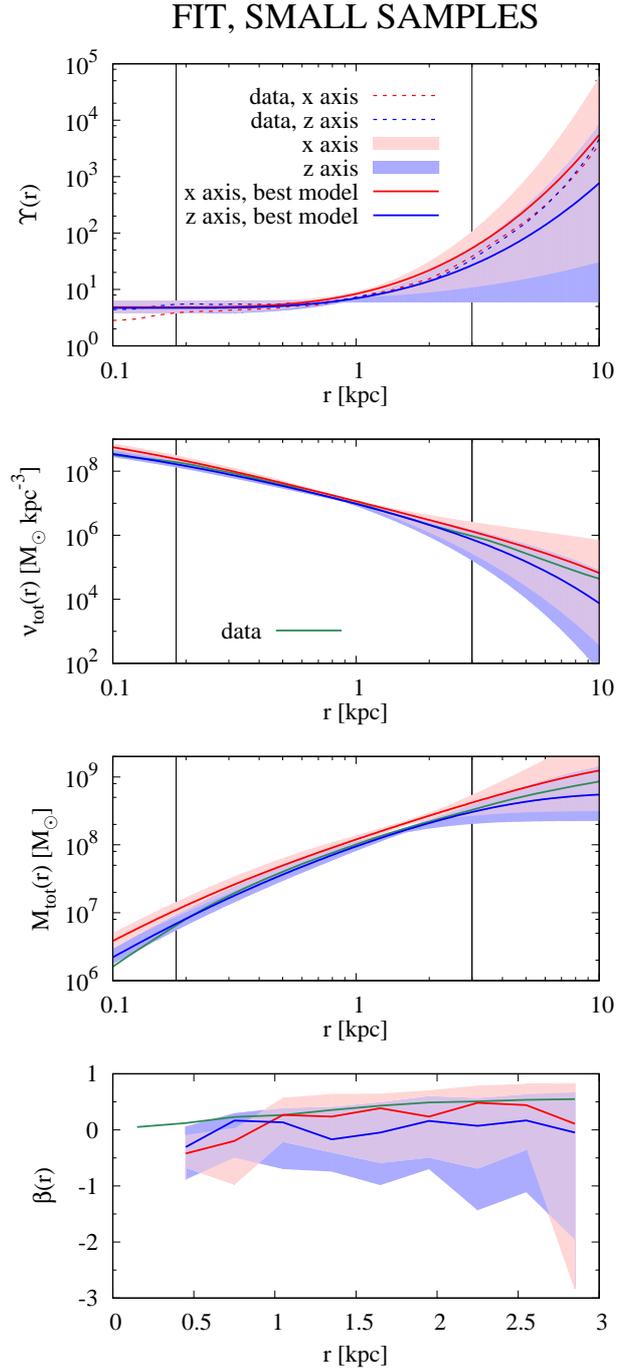}
\caption{Same as Fig.\,\ref{fig:Upsilon_fit} but for two small data samples, one for each line of sight.}
\label{fig:Upsilon_ss}
\end{center}
\end{figure}

The resulting profiles of mass-to-light ratio, total density, total mass and anisotropy are presented in consecutive
panels of Fig.\,\ref{fig:Upsilon_ss} with the solid red (for the observations along the longest axis) and blue
(for the shortest) lines. Shaded areas of corresponding colours show the spread of values of a given parameter
among the mass-to-light ratio models obtained within 1\,$\sigma$ level (innermost loops in Fig.\,\ref{fig:chi_10}).
Dashed (top panel only) and green lines (otherwise) indicate the real values from the simulation for comparison.
Thin vertical lines indicate 3 softening scales for dark matter particles and the outer radius of data sets from
left to right, respectively.

Since the parameters of the mass-to-light ratio profile are strongly degenerated, the obtained density and mass profiles
are satisfactorily well constrained up to the outer boundary of the data sets ($r_{out}=3$\,kpc) and have the same
bias features as for the large data samples: the mass is overestimated at all scales for the observations along the
longest axis whereas for the shortest axis it is well recovered in the inner part of the galaxy but underestimated in
the outer.

Also the anisotropy profiles are similar to those obtained in Sec.\,\ref{density_30}, however for both lines of sight
1\,$\sigma$ errors are much larger and at most radii include the values of the real anisotropy. Overall, the range of
derived anisotropy for the short axis observations tends to lower values, i.e. more tangential orbits are
obtained in this case. Only for the observations along the longest axis the anisotropy can be significantly
overestimated as shown by a larger extension of the light red region towards more radial anisotropy values.

Although for small samples the real values of anisotropy are recovered within 1\,$\sigma$ confidence
level, in contrast to large samples, it does not imply that in this case the anisotropy can be recovered more
accurately. In fact, rather the opposite is true: the input data come with larger sampling errors so the results for
small samples are less precise which allowed for the 1\,$\sigma$ confidence level to contain the real values.

\section{Summary and discussion}
\label{summary}

We studied the systematic errors in the recovered total mass and anisotropy profiles caused by the
ellipticity of a dSph galaxy when using spherically symmetric Schwarzschild modelling method. Although ultimately models
with less symmetry are expected to work better they introduce additional parameters that we believe cannot be constrained
with the presently available data.

For the purpose of the tests we ran a simulation of a major merger of two identical dwarf galaxies, initially composed
of an exponential stellar disc and an NFW-like dark matter halo. The stellar component of the merger remnant had an
axisymmetric prolate shape with the ratio of the shortest to longest principal axis $c/a=0.84$ (or the ellipticity
$\epsilon=1-c/a=0.16$). By observing our remnant along the shortest and longest axis we obtained two datasets
representing the extreme cases of lines of sight which allowed us to measure the maximum bias in the recovered
quantities.

We focused on the determination of the maximum, instead of the mean bias as we find it more informative. Investigating
the `worst-case scenarios', one may expect the results for any other line of sight to lie in between whereas averaging
over many lines of sight would rather prove (or disprove) the reliability of a method for spherical objects. Since in
\citet{kowalczyk_2017} we have shown that the applied method works well for spherically symmetric distribution of
stars, the results presented here are particularly interesting as even for the large data samples the anisotropy is
systematically underestimated and mostly consistent between the extreme lines of sight, however with different
uncertainties (compare the last panel of Fig.\,\ref{fig:Upsilon_fit} and Fig.\,5 in our previous work). Therefore, we
conclude that in a spheroidal galaxy for which the mass profile is unknown, independently of the amount of available
data the anisotropy is not overestimated.

We modelled the total mass content with the mass-to-light ratio varying with radius from the centre of the galaxy
like $\log\Upsilon(r)\sim \big(\log(r)\big)^3$. Such an approach is more convenient than the usual
assumption of two independent components for at least two reasons. First, stars and a dark matter halo are modelled
together so that any possible errors in the deprojection of the stellar distribution can be compensated. Second, the
formula does not impose any particular form of the dark matter density profile, therefore comparisons between different
profiles introduced in the literature (\citealt{einasto_1965}, \citealt{hernquist_1990}, \citealt{burkert_1995},
\citealt{NFW_1997}) are not necessary. Additionally, the dark matter halo is naturally cut off (by the exponential drop
in the stellar distribution) which is not the case e.g. for the NFW profile.  Although it is customary to insert a
cut-off by hand, a scale and functional form of it may influence the modelling and introduce more free parameters if an
orbit library reaches the radii of that sharp drop in dark matter density.

Following the approach already applied in \citet{kowalczyk_2017}, we
focused on two types of data samples which we labelled `large' (all stellar particles from the simulation, i.e.
$4\times 10^5$ particles) and `small' ($1\times 10^5$ particles with positions projected along the line of sight
and 2\,500 line-of-sight velocity measurements). For each type of data we divided the study into additional two steps:
modelling the mock data under the assumption of the known total density profile to derive the anisotropy profile
only and recovering both the anisotropy and the density profile by comparing the absolute values of the objective
function.

Our results show that:
\begin{enumerate}
 \item modelling the galaxy under the assumption that `mass follows light', i.e. that the spatial distribution of the
 total mass can be expressed by rescaling the distribution of the visible tracer, leads to a severe inaccuracy of the
 resulting mass profile (overestimated in the inner parts of the galaxy and underestimated in the outskirts) and
 underestimation of the anisotropy parameter at all scales regardless of the line of sight;

 \item if we assume that the true mass-to-light ratio profile is known the anisotropy is slightly underestimated in the centre
 and well recovered at larger radii for observations along the longest axis, with the accuracy similar to
 the spherical cases which we studied before, whereas for the shortest axis the
 profile is consistent with the one constant with radius and close to $\beta=0$ with sharp drop at the outskirts of the
 galaxy;

 \item when the mass-to-light profile is to be derived, for the observations along the longest axis the total
 mass is overestimated (up to the outer radius of the data set) and the anisotropy is underestimated, however the
 general growing shape of the anisotropy profile is reproduced; for the shortest axis the mass profile is well recovered
 but the
 anisotropy is constant or decreasing with radius; for both lines of sight the mass content at large radii is recovered
 within 1\,$\sigma$ confidence level, therefore, the method seems to be sensitive enough to determine the existence of
 the extended dark matter halo even if the outskirts of the galaxy do not enter the modelling;

 \item when considering small data samples and the true mass-to-light ratio profiles, the mean values of the derived
 anisotropy averaged over many different random samples show the same trends as the results obtained for the large
 samples; sampling errors of the derived anisotropy are $\sim 3$ times larger than sampling errors of the
 real anisotropy;

 \item the derivation of the mass-to-light ratio profiles for two small samples confirms the results obtained for the
 large ones, however the uncertainties are larger.
\end{enumerate}

In summary, for prolate dSph galaxies (expected to be the most typical shape based on the currently preferred formation
models) the determination of velocity anisotropy and total mass depends quite strongly on the viewing angle. If the
projected shape is circular the Schwarzschild method yields an overestimate of the total mass. If the projected shape
of the dSph is elongated the mass is well recovered. In both cases the anisotropy is generally underestimated, but
more so in the latter. This is understandable as in a prolate spheroidal system a large fraction of the orbits are
elongated along the long axis and the system has the largest velocity dispersion along this axis (see
Fig.\,\ref{fig:maps_all}).

Studies of dwarf galaxies in the Local Group based on Jeans modelling often result in close to zero or negative values
of anisotropy assumed to be constant with radius (\citealt{lokas_2005}, \citealt{lokas_2009}, \citealt{walker_2009}).
Our work shows that it is not unexpected. The simultaneous recovering of the mass and anisotropy profiles results in
the flat anisotropy profiles with negative mean values.

\citet{jardel_2012} who modelled the Fornax dSph with axisymmetric Schwarzschild method based on the full line-of-sight
velocity profile (while we used only moments), though assuming the orientation
of the galaxy, reported opposite results. The Fornax dSph agrees well with our mock data as it is elongated, shows
traces of a major merger about 6\,Gyr ago \citep{delpino_2015} with no strong interaction with Milky Way due to its
extended orbit \citep{battaglia_2015} and its data sample is similar in size.
\citet{jardel_2012} recovered the profile of $\sigma_r/\sigma_t=(\beta-1)^{-1/2}$ which is mostly constant, close to
1 in the centre and rises to 1.5 at larger radii ($\beta$ rising from 0 to 0.5). Such a profile is consistent with our
findings for the true mass-to-light ratio profile and observations along the longest axis. It may mean that their
treatment of the elliptical projected shape of the galaxy lifted the bias. However, they also fitted the dark matter
halo profile. In this case for small samples we were able to recover the true profile of anisotropy within 1\,$\sigma$
confidence level but our uncertainties were much larger than shown by \citet{jardel_2012}.

In contrast to other authors applying the Schwarzschild method to dwarf galaxies (\citealt{jardel_2012},
\citealt{jardel_2013}, \citealt{breddels_2013}, \citealt{breddels_2013b}) we did not attempt to recover the inner
profile of the dark matter halo. The inner profile is poorly constrained as it affects only the kinematics of stars in
the very centre of the galaxy, ultimately requiring large amount of data in this region. Unfortunately, current
observational data do not seem sufficient for this purpose. Our realistic small data samples presented in
Sec.\,\ref{small_sample} cover a potential dark matter core with only 1-2 data points with large error bars. Therefore,
with such a general method as Schwarzschild modelling deviations caused by a dark matter core would be comparable to
uncertainties. However, a test of the ability of the method to recover the inner slope using large data samples might
be interesting and give some insight on future developments in solving the `cusp-core' problem.

We conclude that the spherical Schwarzschild modelling method proves to be useful also when applied to non-idealized,
spheroidal objects created by a collision of galaxies. It is able to provide us with good estimates of the mass profile
also at large radii and (at least) the lower limit on the anisotropy.

\section*{Acknowledgements}

This research was supported in part by the
Polish Ministry of Science and Higher Education under grant 0149/DIA/2013/42 within the Diamond Grant Programme for
years 2013-2017 and by the Polish National Science Centre under grant 2013/10/A/ST9/00023. MV acknowledges support
from HST-AR-13890.001, NSF awards AST-0908346, AST-1515001, NASA-ATP award NNX15AK79G.

\bsp
\label{lastpage}

\begin{thebibliography}{}
\bibitem[\protect\citeauthoryear{Battaglia, Sollima \& Nipoti}{Battaglia
 et al.}{2015}]{battaglia_2015}
Battaglia G., Sollima A., Nipoti C., 2015, \mnras, 454, 2401

\bibitem[\protect\citeauthoryear{Binney \& Mamon}{Binney \&
  Mamon}{1982}]{binney_1982}
Binney J.,  Mamon G.~A.,  1982, \mnras, 200, 361

\bibitem[\protect\citeauthoryear{Binney \& Tremaine}{Binney \&
  Tremaine}{2008}]{GD}
Binney J.,  Tremaine S.,  2008, Galactic Dynamics, 2 edn.
Princeton University Press, Princeton, NJ

\bibitem[\protect\citeauthoryear{Breddels \& Helmi}{Breddels \&
  Helmi}{2013}]{breddels_2013b}
Breddels M.~A.,  Helmi A.,  2013, \aap, 558, A35

\bibitem[\protect\citeauthoryear{Breddels, Helmi, van~den Bosch, van~de Ven  \&
  Battaglia}{Breddels et~al.}{2013}]{breddels_2013}
Breddels M.~A.,  Helmi A.,  van~den Bosch R. C.~E.,  van~de Ven G.,   Battaglia
  G.,  2013, \mnras, 433, 3173

\bibitem[\protect\citeauthoryear{Burkert}{Burkert}{1995}]{burkert_1995}
Burkert A.,  1995, \apj, 447, L25

\bibitem[\protect\citeauthoryear{Campbell et~al.,}{Campbell
  et~al.}{2017}]{campbell_2017}
Campbell D.~J.~R. et~al., 2017, \mnras, 469, 2335

\bibitem[\protect\citeauthoryear{Cappellari et~al.,}{Cappellari
  et~al.}{2006}]{cappellari_2006}
Cappellari M.  et~al., 2006, \mnras, 366, 1126

\bibitem[\protect\citeauthoryear{Cretton \& Emsellem}{Cretton \&
  Emsellem}{2004}]{cretton_2004}
Cretton N.,  Emsellem E.,  2004, \mnras, 327, L31

\bibitem[\protect\citeauthoryear{Cretton, van~den Bosch \& Frank}{Cretton
  et~al.}{1999}]{cretton_1999}
Cretton N., {\lowercase{V}an}~den Bosch R. C.~E., Frank C., 1999, \apj, 514, 704

\bibitem[\protect\citeauthoryear{Ebrov\'a \& {\L}okas}{Ebrov\'a \&
{\L}okas}{2015}]{ebrova_2015}
Ebrov\'a I., {\L}okas E.~L., 2015, \mnras, 394, L102

\bibitem[\protect\citeauthoryear{Einasto}{Einasto}{1965}]{einasto_1965}
Einasto J., 1965, Trudy Astrofizicheskogo Instituta Alma-Ata, 5, 87

\bibitem[\protect\citeauthoryear{Gebhardt et~al.,}{Gebhardt
  et~al.}{2003}]{gebhardt_2003}
Gebhardt K.  et~al., 2003, \apj, 583, 92

\bibitem[\protect\citeauthoryear{Gilmore, Wilkinson, Wyse, Kleyna, Koch  \&
  Evans}{Gilmore et~al.}{2007}]{gilmore_2007}
Gilmore G.,  Wilkinson M.~I.,  Wyse R. F.~G.,  Kleyna J.~T.,  Koch A.,   Evans
  N.~W.,  2007, \apj, 663, 948

\bibitem[\protect\citeauthoryear{Governato et~al.,}{Governato
  et~al.}{2010}]{governato_2010}
Governato F.  et~al., 2010, \nat, 463, 203

\bibitem[\protect\citeauthoryear{Hernquist}{Hernquist}{1990}]{hernquist_1990}
Hernquist L.,  1990, \apj, 356, 359


\bibitem[\protect\citeauthoryear{Jardel \& Gebhardt}{Jardel \&
  Gebhardt}{2012}]{jardel_2012}
Jardel J.~R.,  Gebhardt K.,  2012, \apj, 746, 89

\bibitem[\protect\citeauthoryear{Jardel, Gebhardt, Fabricius, Drory  \&
  Williams}{Jardel et~al.}{2013}]{jardel_2013}
Jardel J.~R.,  Gebhardt K.,  Fabricius M.~H.,  Drory N.,   Williams M.~J.,
  2013, \apj, 763, 91

\bibitem[\protect\citeauthoryear{Kazantzidis, {\L}okas, Callegari, Mayer \&
Moustakas}{Kazantzidis et~al.}{2011a}]{kazantzidis_2011a}
Kazantzidis S.,  {\L}okas E.~L.,  Callegari S., Mayer L., Moustakas L.~A.,
2011a, \apj, 726, 98

\bibitem[\protect\citeauthoryear{Kazantzidis, {\L}okas, Mayer, Knebe \&
Klimentowski}{Kazantzidis et~al.}{2011b}]{kazantzidis_2011b}
Kazantzidis S.,  {\L}okas E.~L.,   Mayer L., Knebe A., Klimentowski J.,
 2011b, \apj, 740, L24

\bibitem[\protect\citeauthoryear{Kazantzidis, {\L}okas, Mayer}{Kazantzidis et~al.}
{2013}]{kazantzidis_2013}
Kazantzidis S.,  {\L}okas E.~L., Mayer L., 2013, \apj, 764, L29

\bibitem[\protect\citeauthoryear{Kowalczyk, {\L}okas, Kazantzidis \& Mayer}{Kowalczyk
 et al.}{2013}]{kowalczyk_2013}
Kowalczyk K., {\L}okas E.~L., Kazantzidis S., Mayer L., 2013, \mnras, 431, 2796

\bibitem[\protect\citeauthoryear{Kowalczyk, {\L}okas  \& Valluri}{Kowalczyk
  et~al.}{2017}]{kowalczyk_2017}
Kowalczyk K.,  {\L}okas E.~L.,   Valluri M.,  2017, \mnras, 470, 3959

\bibitem[\protect\citeauthoryear{Lima, Gerbal \& M\'arquez}
{Lima, Gerbal \& M\'arquez}{1999}]{lima_1999}
Lima Neto G.~B., Gerbal D., M\'arquez I.,  1999, \mnras, 309, 481

\bibitem[\protect\citeauthoryear{{\L}okas}{{\L}okas}
{2002}]{lokas_2002}
{\L}okas E.~L., 2002, \mnras, 333, 697

\bibitem[\protect\citeauthoryear{{\L}okas}{{\L}okas}
{2009}]{lokas_2009}
{\L}okas E.~L., 2009, \mnras, 394, L102

\bibitem[\protect\citeauthoryear{{\L}okas \& Mamon}{{\L}okas \&
  Mamon}{2003}]{lokas_2003}
{\L}okas E.~L.,  Mamon G.~A.,  2003, \mnras, 343, 401

\bibitem[\protect\citeauthoryear{{\L}okas, Mamon \& Prada}{{\L}okas
 et al.}{2005}]{lokas_2005}
{\L}okas E.~L.,  Mamon G.~A., Prada F., 2005, \mnras, 363, 918

\bibitem[\protect\citeauthoryear{{\L}okas, Majewski, Kazantzidis, Mayer, Carlin, Nidever \&
Moustakas}{{\L}okas et~al.}{2012}]{lokas_2012}
{\L}okas E.~L., Majewski S. R., Kazantzidis S., Mayer L., Carlin J. L., Nidever D. L., Moustakas L.~A.,
2012, \apj, 751, 61

\bibitem[\protect\citeauthoryear{{\L}okas, Gajda \& Kazantzidis}{{\L}okas
 et al.}{2013}]{lokas_2013}
{\L}okas E.~L.,  Gajda G., Kazantzidis S., 2013, \mnras, 433, 878

\bibitem[\protect\citeauthoryear{{\L}okas, et al.}{{\L}okas
 et al.}{2014}]{lokas_2014}
{\L}okas E.~L.,  Ebrov\'a I., del Pino A., Semczuk M., 2014, \mnras, 445, L6

\bibitem[\protect\citeauthoryear{Mateo}{Mateo}{1998}]{mateo_1998}
Mateo M.,  1998, \araa, 36, 435

\bibitem[\protect\citeauthoryear{McConnachie}{McConnachie}{2012}]{mcconnachie_2012}
McConnachie A.~W.,  2012, \aj, 144, 4

\bibitem[\protect\citeauthoryear{Navarro, Frenk  \& White}{Navarro
  et~al.}{1997}]{NFW_1997}
Navarro J.~F.,  Frenk C.~S.,   White S. D.~M.,  1997, \apj, 490, 493

\bibitem[\protect\citeauthoryear{\lowercase{Del} Pino, Aparicio \& Hidalgo}
{\lowercase{Del} Pino et al.}{2015}]{delpino_2015}
\lowercase{Del} Pino A., Aparicio A., Hidalgo S.~L.,  2015, \mnras, 454, 3996

\bibitem[\protect\citeauthoryear{Press, Teukolsky, Vetterling  \&
  Flannery}{Press et~al.}{1992}]{NR}
Press W.~H.,  Teukolsky S.~A.,  Vetterling W.~T.,   Flannery B.~P.,  1992,
  Numerical Recipes in C, 2 edn.
Cambridge University Press, New York, NY

\bibitem[\protect\citeauthoryear{Rix, de Zeeuw, Cretton, van~der Marel  \&
  Carollo}{Rix et~al.}{1997}]{rix_1997}
Rix H.-W.,  de Zeeuw P.~T.,  Cretton N.,  van~der Marel R.~P.,   Carollo C.~M.,
   1997, \apj, 488, 702

\bibitem[\protect\citeauthoryear{Sawala et al.}{Sawala et al.}{2016}]{sawala_2016}
Sawala T. et~al., 2016, \mnras, 457, 1931

\bibitem[\protect\citeauthoryear{Schwarzschild}{Schwarzschild}{1979}]{schwarzschild_1979}
Schwarzschild M.,  1979, \apj, 232, 236

\bibitem[\protect\citeauthoryear{S\'ersic}{S\'ersic}{1968}]{sersic_1968}
S\'ersic J.~L., 1968, Atlas de galaxias australes, Observatorio Astronomico,
Cordoba


\bibitem[\protect\citeauthoryear{Springel}{Springel}{2005}]{springel_2005}
Springel V.,  2005, \mnras, 364, 1105

\bibitem[\protect\citeauthoryear{{The CGAL Project}}{{The CGAL
  Project}}{2015}]{cgal}
{The CGAL Project} 2015, {CGAL} User and Reference Manual, {4.7} edn.
{CGAL Editorial Board}, \url {http://doc.cgal.org/4.7/Manual/packages.html}

\bibitem[\protect\citeauthoryear{Thomas, Saglia, Bender, Thomas, Gebhardt,
  Magorrian  \& Richstone}{Thomas et~al.}{2004}]{thomas_2004}
Thomas J.,  Saglia R.~P.,  Bender R.,  Thomas D.,  Gebhardt K.,  Magorrian J.,
   Richstone D.,  2004, \mnras, 353, 391

\bibitem[\protect\citeauthoryear{Valluri, Merritt  \& Emsellem}{Valluri
  et~al.}{2004}]{valluri_2004}
Valluri M.,  Merritt D.,   Emsellem E.,  2004, \apj, 602, 66

\bibitem[\protect\citeauthoryear{{\lowercase{V}an}~den Bosch \& de
  Zeeuw}{{\lowercase{V}an}~den Bosch \& de~Zeeuw}{2010}]{vdBosch_2010}
{\lowercase{V}an}~den Bosch R. C.~E.,  de Zeeuw P.~T.,  2010, \mnras, 401, 1770

\bibitem[\protect\citeauthoryear{{\lowercase{V}an}~der Marel, Cretton, de Zeeuw
   \& Rix}{{\lowercase{V}an}~der Marel et~al.}{1998}]{vdMarel_1998}
{\lowercase{V}an}~der Marel R.~P.,  Cretton N.,  de Zeeuw P.~T.,   Rix H.-W.,
  1998, \apj, 493, 613

\bibitem[\protect\citeauthoryear{Walker et al.}{Walker et al.}
{2009}]{walker_2009}
Walker M.~G., Mateo M., Olszewski E.~W., Pa\~narrubia J., Evans N.~W.,
Gilmore G.,  2009, \apj, 704, 1274

\bibitem[\protect\citeauthoryear{Widrow \& Dubinski}{Widrow \&
  Dubinski}{2005}]{widrow_2005}
Widrow L.~M., Dubinski J.,  2005, \apj, 631, 838

\bibitem[\protect\citeauthoryear{Widrow, Pym \& Dubinski}{Widrow
 et~al.}{2008}]{widrow_2008}
Widrow L.~M., Pym B., Dubinski J.,  2008, \apj, 679, 1239

\bibitem[\protect\citeauthoryear{Wolf, Martinez, Bullock, Kaplinghat, Geha,
  Mu\~noz, Simon  \& Avedo}{Wolf et~al.}{2010}]{wolf_2010}
Wolf J.,  Martinez G.~D.,  Bullock J.~S.,  Kaplinghat M.,  Geha M.,  Mu\~noz
  R.~R.,  Simon J.~D.,   Avedo F.~F.,  2010, \mnras, 406, 1220
\end{thebibliography}
\end{document}